\documentclass[preprint2,10.5pt]{aastex}
\begin{document}

\title{\large{\rm{NEW EVIDENCE SUPPORTING MEMBERSHIP FOR TW NOR IN LYNG{\AA} 6 AND THE CENTAURUS SPIRAL ARM\thanks{Based on observations gathered with the ESO-VISTA telescope (proposal ID 172.B-2002).}}}}
\author{ D.~Majaess$^{1,}$\altaffilmark{*}, D.~Turner$^{1}$, C. Moni Bidin$^{2}$, F. Mauro$^{2}$, D. Geisler$^{2}$, W. Gieren$^{2}$, D. Minniti$^{3,4,5}$, A-N. Chen\'e$^{2,7}$, P. Lucas$^{6}$, J. Borissova$^{7}$, R. Kurtev$^{7}$, I. D\'ek\'any$^{3}$, R. Saito$^{3}$} 
\affil{$^{1}$Saint Mary's University, Halifax, Nova Scotia, Canada}
\affil{$^{2}$Universidad de Concepci\'on, Concepci\'on, Chile}
\affil{$^{3}$Pontificia Universidad Catolica de Chile, Santiago, Chile} 
\affil{$^{4}$Vatican Observatory, Vatican City, Italy}
\affil{$^{5}$Princeton University, Princeton, New Jersey, USA}
\affil{$^{6}$University of Hertfordshire, Hatfield, United Kingdom}
\affil{$^{7}$Universidad de Valpara\'iso, Valpara\'iso, Chile}
\altaffiltext{*}{dmajaess@cygnus.smu.ca}
 
\begin{abstract}
The putative association between the $10.8^{\rm{d}}$ classical Cepheid TW Nor and the open cluster Lyng{\aa} 6 has generated considerable debate in the literature. New {\it JHK}$_s$ photometry in tandem with existing radial velocities for Lyng{\aa} 6 stars imply cluster membership for TW Nor, and establish the variable as a high-weight calibrator for classical Cepheid relations.  Fundamental mean parameters determined for Lyng{\aa} 6 are: $d=1.91\pm0.10$ kpc, $E(J-H)=0.38\pm0.02$, and $\log{\tau}=7.9\pm0.1$.  The \citet{be07}/\citet{tu10} Galactic {\it VI$_c$} Wesenheit function was revised using TW Nor's new parameters: {\it W}$_{VI,0}=(-3.37\pm0.08) \log{P_0}-2.48\pm0.08$.  TW Nor/Lyng{\aa} 6 lie beyond the Sagittarius-Carina spiral arm and occupy the Centaurus arm, along with innumerable young Cepheids and clusters (e.g., VW Cen \& VVV CL070).
\end{abstract}


\section{{\rm \footnotesize INTRODUCTION}}
Cluster membership provides a means to constrain the distance, color excess, and age of a Cepheid \citep{tu92}.  The parameters are subsequently employed to calibrate Cepheid period-luminosity, period-Wesenheit, and period-age relationships \citep{tu10}.  Such functions bolster efforts to map the Galaxy's local spiral structure \citep[][and references therein]{ma09}, establish the extragalactic distance scale \citep{pi04,mr09}, and determine the Hubble constant \citep{fm10}.  Establishing a connection between the $10.8^{\rm{d}}$  classical Cepheid TW Nor and open cluster Lyng{\aa} 6 is consequently desirable.

\citet{ts66} first noted that TW Nor may be a member of Lyng{\aa} 6.  However, \citet{mv75} doubted the cluster's existence after evaluating {\it UBV-H}$\beta$ photometry for 18 stars.   \citet{ma75} obtained {\it UBV} photoelectric photometry for 8 stars and argued for cluster parameters of $d\simeq2.5$ kpc and {\it E(B-V)}$=1.37\pm0.02$.  \citet{ma75} remarked that the luminosity implied for TW Nor from cluster membership matched expectations from period-luminosity calibrations.  \citet{vh76} provided {\it BV} photoelectric and photographic photometry for 38 stars, and revised the cluster distance downward ($d\simeq1.6$ kpc).   \citet{dp76} concluded that evolutionary models yielded similar ages for TW Nor and Lyng{\aa} 6 (40 Myr), thereby supporting cluster membership for the Cepheid.  \citet{ly77} supplemented existing data with $uvby \beta$ photometry and temperature classifications (objective prism spectra) for several stars, and corroborated the parameters cited for Lyng{\aa} 6 by \citet{vh76}.  

Uncertainties tied to TW Nor's reddening \citep{vh76,fm83} and distance complicated efforts to calibrate classical Cepheid relations throughout the 1980s.  For example, \citet{mc83} included TW Nor in their calibration after adopting a mean of the \citet{ma75} and \citet{vh76} distances to Lyng{\aa} 6.

\begin{figure}[!t]
\epsscale{1}
\plotone{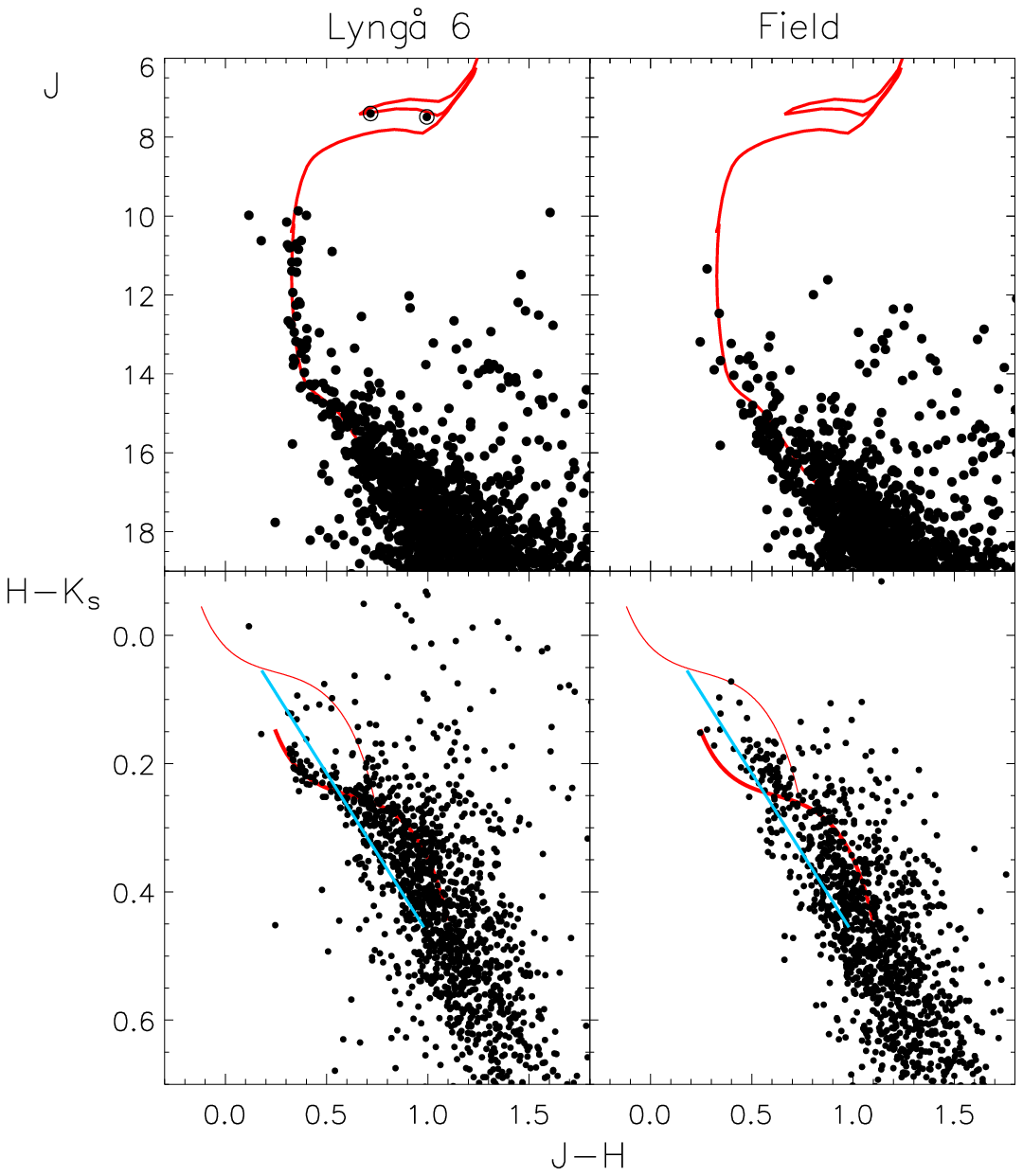}
\caption{{\small VVV color-magnitude and color-color diagrams for Lyng{\aa} 6 stars and a comparison field.  A population of mid-to-late B-type cluster stars is absent from the latter.  Stars left of the cyan line are typically earlier than $\simeq$F5 \citep{sl09,tu11}.  The isochrone and intrinsic fit were adjusted (reddened) to match the observed data.  Ly6-4 \citep[an evolved red star,][]{ly77} and the classical Cepheid TW Nor (circled dots) share a common radial velocity \citep{me87,me08}.  Applying the intrinsic color-color relation of \citet{tu11} and a $\log{\tau}=7.9\pm0.1$ Padova isochrone to Lyng{\aa} 6 stars yields:{\it E(J-H)}$=0.38\pm0.02$ and $d=1.91\pm0.10$ kpc. The extinction laws adopted are described in the text.}}
\label{fig-cm}
\end{figure}

\citet{wa85} acquired deep {\it BVI}$_c$ CCD observations that extended existing photometry to $V\simeq 20$, and determined cluster parameters of $d=2.0\pm0.2$ kpc and $\tau \simeq 10^8$ years.  The age is older than that estimated for TW Nor and Lyng{\aa} 6 by \citet{dp76}.  An independent cluster reddening could not be established by \citet{wa85} since the B-star sequence is parallel to the reddening vector in {\it B-V/V-I}$_c$ color-color diagrams. \citet{wa85} adopted the \citet{vh76} reddening to evaluate the distance to Lyng{\aa} 6.  The \citet{db94} pulsational parallax for TW Nor disagreed with the cluster distance adopted, and {\it UBVRI} polarization measurements by \citet{or01} exhibited a $\ge2 \sigma$ mismatch between cluster members and TW Nor.  \citet{ho03} supplied new {\it UBV} CCD photometry and established $d\simeq1.7$ kpc and {\it E(B-V)}$=1.36\pm0.08$ for Lyng{\aa} 6.  By contrast, \citet{lc07} advocated a smaller reddening for the Cepheid of {\it E(B-V)}$=1.17$, and \citet{ka09} determined a closer distance to Lyng{\aa} 6 ($d=1.36$ kpc) based on $uvby \beta$ photometry for five stars.  The latter result negates cluster membership for TW Nor.

In summary, parameters for TW Nor/Lyng{\aa} 6 and their putative association have been debated for nearly a half century.  In this study, new {\it JHK}$_s$ VVV (ESO Public Survey `VISTA Variables in the V\'ia L\'actea')  photometry is presented which substantiates the cluster's existence, bolsters cluster membership for TW Nor, and provides a reliable set of fundamental parameters to allow for the subsequent calibration of classical Cepheid relations. 
 
\section{{\rm \footnotesize VVV PHOTOMETRY}}
The VVV survey aims to establish precise multi-epoch photometry for fields in the Galactic bulge and near the Galactic plane \citep[$\ell=295-10 \degr$,][]{mi10}.  VVV images extend magnitudes fainter and exhibit increased spatial resolution versus 2MASS, which is particularly important for mitigating contaminated photometry in crowded regions near the Galactic center and the cores of globular clusters \citep[M15,][]{ma10c}. The VVV survey will provide precise multi-epoch photometry for stars and variables in  $\sim 3 \times 10^2$ open clusters and $\ge39$ globular clusters (e.g., NGC 6441).

Details of the pipeline constructed to process and extract VVV photometry are discussed in Mauro et al. (2011, in prep.).  PSF photometry performed using DAOPHOT was subsequently tied to 2MASS {\it JHK}$_s$ standards \citep[see][]{mb11}. 

\section{{\rm \footnotesize ANALYSIS}}
A color-color diagram was compiled for Lyng{\aa} 6 stars within $\simeq 2 \arcmin$ of J2000 coordinates 16:04:54.42 -51:57:31.6 (Fig.~\ref{fig-cm}).  A diagram was likewise constructed for a comparison field 20$\arcmin$ adjacent to Lyng{\aa} 6.  Reddening laws characterizing dust along the line of sight were derived to evaluate the color excess.  The reddening vector may be determined by tracking deviations of red clump stars from their mean intrinsic color owing to extinction.  The mean intrinsic color was inferred from nearby red clump stars ($d\le50$ pc) with revised Hipparcos parallaxes \citep{vl07}.  Reddening corrections were neglected since the calibrating sample lies in close proximity to the Sun.  Spurious data were removed, and the resulting mean parameters for red clump stars are highlighted in Table (\ref{table1}).  The values correspond to a K0III star as inferred from catalogued\footnote{Catalogue of Stellar Spectral Classifications (B. Skiff), http://vizier.cfa.harvard.edu/viz-bin/VizieR?-source=B/mk} spectroscopic classifications.  A reddening vector of $E(J-H)/E(H-K_s)=1.94\pm0.03$ was determined, and is supported by the work of \citet{sl08}.  

Applying the reddening vector to the Lyng{\aa} 6 color-color diagram reveals a prominent sequence of mid-to-late B stars with {\it E(J-H)}$=0.38\pm0.03$ (Fig.~\ref{fig-cm}).  The corresponding value of $E(B-V)$ is sensitive to the conversion from $E(J-H)$ adopted \citep[see][and references therein]{ma08}, but lies within the range of 1.13 to 1.40 mag.  A visible sequence of young cluster stars is absent from the comparison field.   Lyng{\aa} 6 stars terminate near B5 according to the intrinsic {\it JHK}$_s$ relations of \citet{sl09} and \citet{tu11}.   

\begin{deluxetable}{cccccc}
\tablewidth{0pt}
\tabletypesize{\small}
\tablecaption{{\small Intrinsic Parameters for Red Clump Stars}}
\tablehead{\colhead{$M_J$} & \colhead{$M_H$} & \colhead{$M_{K_s}$} & \colhead{$(J-H)_0$} & \colhead{$(H-K_s)_0$} & \colhead{$(J-K_s)_0$ }}
\startdata
 $-1.04\pm0.04$ & $-1.52\pm0.04$ & $-1.59\pm0.04$ & $0.48\pm0.06$ & $0.07\pm0.06$ & $0.55\pm0.06$ 
\enddata
\label{table1}
\end{deluxetable}

Color-magnitude diagrams for Lynga 6 and an adjacent comparison field are shown in Fig.~\ref{fig-cm}.  The population of cluster stars is absent from the comparison field.  A $\log(\tau)=7.9\pm0.1$ Padova isochrone \citep{bo04} was adopted based on the reddening and spectral types inferred from the color-color diagram. That age provides an evolutionary track which aptly matches bluer cluster stars, an evolved red star at J2000 coordinates 16:04:54.42 -51:57:32.5 \citep[hereafter Ly6-4,][]{ly77}, and TW Nor.  

Radial velocity measurements from \citet{me87,me08} for Ly6-4 agree with that established for TW Nor  ($RV\simeq-56$ km/s).  The radial velocity ties TW Nor to that star, and both are surrounded by bluer cluster members that adhere to the expected evolutionary track.  The pair lie near the cluster core and are separated by $22 \arcsec$.  The stars are brighter than the saturation limit of the VVV survey and thus the {\it JHK}$_s$ photometry was taken from 2MASS.  The color-magnitude and color-color diagrams compiled from VVV photometry establishes Lyng{\aa} 6 as a {\it bona fide} open cluster (Fig.~\ref{fig-cm}).   

\citet{or01} concluded that a $2 \sigma$ mismatch exists between polarization measurements for cluster members and TW Nor.  {\it JHK}$_s$ photometry for the \citet{or01} sample confirm that the polarization measurements efficiently segregate cluster members from field stars.  Yet polarization measurements imply cluster membership for Ly6-4 (the evolved red star, Fig~\ref{fig-cm}). A contradiction consequently emerges since CORAVEL measurements by \citet{me87,me08} link that star and TW Nor to a common radial velocity.  The pair lie at the cluster's core and are separated by $22 \arcsec$. Incidentally, 2MASS {\it JHK}$_s$ colors for Ly6-4 are inconsistent with a G-type classification \citep{ly77}, for the cluster's reddening.  The anomalous {\it JHK}$_s$ (\& UBV) colors may be tied to the star's binary nature \citep{me87,me08}.

A ratio of total to selective extinction ({\it R}) for the field was determined to evaluate the distance to Lyng{\aa} 6.  The precision and faintness attained by the VVV survey permits an independent assessment of {\it R} using red clump stars via the variable extinction method \citep{tu76}.  The expression for computing the distance to red clump stars simplifies since mean intrinsic parameters may be adopted:
\begin{eqnarray}
\nonumber
J-M_J-R_J \times ( (J-H)-(J-H)_0)=\mu_0 \\
\nonumber
J=R_J \times (J-H) + {\rm const} 
\end{eqnarray}
A determination of {\it R} follows by correlating {\it J} and {\it J-H} for a sample of red clump stars at a common distance.  A histogram of the color-magnitude diagram reveals a sizable population of red clump stars $\sim5$ kpc distant. Comparing {\it J} and {\it J-H} photometry for that group of red clump stars yields $A_J/E(J-H)=R_J=2.75\pm0.07$.  The extinction ratios for the field follow from $A_J/E(J-H)$ and $E(J-H)/E(H-K_s)$ via:
  \begin{eqnarray}
\nonumber
A_J/A_{H}=\left( A_J/E_{J-H}\right)  / \left( A_J/E_{J-H} -1 \right)   \\
\nonumber
A_J/A_{K_s}=\left( \frac{A_J}{A_H} \frac{E_{J-H}}{E_{H-K_s}} \right) / \left( \frac{E_{J-H}}{E_{H-K_s}} - \frac{A_J}{A_H} +1 \right) \nonumber
\end{eqnarray}

The resulting extinction ratios ($A_J/A_H:A_J/A_{K_s}=1.57:2.22$) agree with the canonical values \citep[e.g.,][]{bo04}. 

\begin{figure}[!t]
\epsscale{0.8}
\plotone{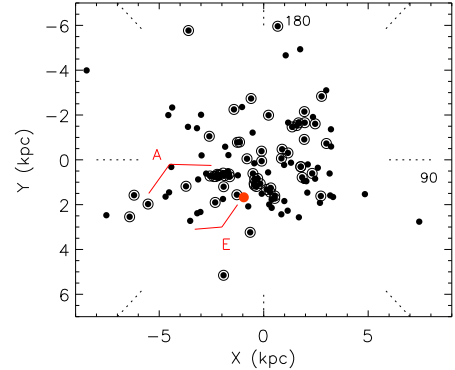}
\caption{{\small A map of local spiral structure as delineated by long period classical Cepheids (dots) and young clusters (circled dots) \citep[see also][]{ma09}.  The Carina (A) and Centaurus (E) spiral arms are indicated on the diagram.  TW Nor/Lyng{\aa} 6 (red dot) reside in the Centaurus spiral arm.}}
\label{fig-gs}
\end{figure}

The final parameters for Lyng{\aa} 6 are $d=1.91\pm0.10$ kpc, ${\it E(J-H)}=0.38\pm0.02$, and $\log{\tau}=7.9\pm0.1$.  The zero-point of the Padova isochrone employed matches a distance scale anchored to seven benchmark clusters featuring equivalent revised Hipparcos and {\it JHK}$_s$ ZAMS distances \citep{ma11}, to within the uncertainties.  The benchmark clusters are the Hyades, $\alpha$ Per, Praesepe, Coma Ber, IC 2391, IC 2609, and NGC 2451 \citep{vl09,ma11}.  The distance scale can (presently) rely on a suite of clusters that are independent of the Pleiades, and where consensus exists regarding the distances \citep{ma11}. Models should be calibrated and evaluated using those seven nearby clusters where consensus exists regarding the distances, rather than the one discrepant cluster (i.e. the Pleiades).

A map of local spiral structure (Fig.~\ref{fig-gs}) illustrates the location of TW Nor/Lyng{\aa} 6 within the broader context of the Milky Way.  Fig.~\ref{fig-gs} employs young open clusters and long period classical Cepheids to trace the Galaxy's spiral structure.  The new distance established here implies that TW Nor lies beyond the Sagittarius-Carina arm and occupies the Centaurus arm (Fig.~\ref{fig-gs}).  That conclusion is supported by the distance determined for TW Nor via the infrared surface brightness technique \citep[][in press]{st11}. The long period classical Cepheids KQ Sco, QY Cen, VW Cen, OO Cen, and KN Cen likewise delineate the Centaurus arm \citep[see also][]{ma09}. The classical Cepheids are listed in order of decreasing Galactic longitude ($\ell$).  The young open clusters VVV CL070 \citep{bo11}, and Hogg 15 occupy the Centaurus arm, etc.  Deep {\it JHK}$_s$ VVV and UKIDSS \citep{lu08} photometry shall provide further constraints on the Galaxy's morphology by facilitating the discovery of Cepheids and star clusters \citep{mi11,mb11}.  

\begin{figure}[!t]
\epsscale{.8}
\plotone{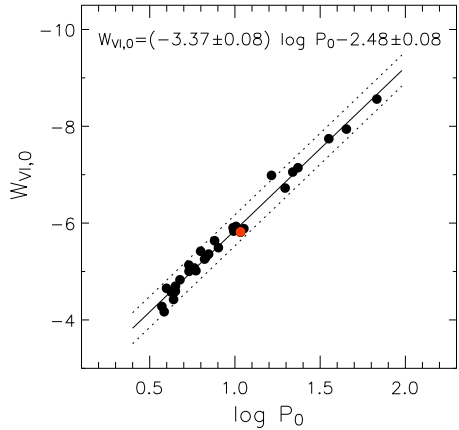}
\caption{{\small The hybrid Galactic {\it VI}$_c$ Wesenheit function for 32 classical Cepheid calibrators \citep{be07,tu10}. The Wesenheit magnitude for TW Nor (red dot) is tied to the revised distance for Lyng{\aa} 6.}}
\label{fig-wf}
\end{figure}

TW Nor may be used as a high-weight calibrator for the universal Wesenheit template \citep{ma10c,ma11} and classical Cepheid relations.  A recent version of of the Galactic {\it VI}$_c$ Wesenheit calibration consists of 10 nearby classical Cepheids with precise HST parallaxes and 24 cluster Cepheids \citep{be07,tu10}, referred to hereafter as the hybrid relation. The new parameters for TW Nor place the Cepheid near the mean Wesenheit trend (Fig.~\ref{fig-wf}). The Wesenheit magnitude for TW Nor ($W_{VI{_c},0}=-5.80$) agrees with that established for calibrators exhibiting similar periods ($\zeta$ Gem \& V340 Nor).  The hybrid Galactic {\it VI}$_c$ Wesenheit function was updated to include the revised parameters for TW Nor: 
\begin{eqnarray}
W_{VI{_c},0}=(-3.37\pm0.08) \log{P_0}-2.48\pm0.08
\label{eqn-w}
\end{eqnarray}
The canonical extinction law was employed ($R_{VI_c}=2.55$). The short period classical Cepheid SU Cas was excluded from the derivation and Fig.~\ref{fig-wf} since its parameters are being revised by Turner \citep[see also][]{st11}.  

\section{{\rm \footnotesize CONCLUSION}}
Color-magnitude and color-color diagrams constructed from precise {\it JHK}$_s$ VVV photometry substantiate the existence of the open cluster Lyng{\aa} 6 (Fig.~\ref{fig-cm}).  The diagrams feature a sequence of mid-to-late B-type stars exhibiting a mean color excess of {\it E(J-H)}$=0.38\pm0.02$ (Fig.~\ref{fig-cm}).  That sequence is absent from an adjacent comparison field.  The brightest cluster members are the classical Cepheid TW Nor and Ly6-4 (an evolved red star, Fig.~\ref{fig-cm}).  The objects share equivalent radial velocities and are surrounded by bluer cluster stars which follow a $\log(\tau)=7.9\pm0.1$  evolutionary track.  The distance for Lyng{\aa} 6 ($d=1.91\pm0.10$ kpc) results after correcting for extinction using reddening laws inferred from red clump stars ($E(J-H)/E(H-K_s)=1.94\pm0.03$).  The result agrees with the distance for TW Nor obtained by \citet[][in press]{st11} via the infrared surface brightness technique, which likewise implies membership for the Cepheid in Lyng{\aa} 6 and the Centaurus spiral arm.  The revised parameters for TW Nor were employed to update the \citet{be07}/\citet{tu10} Galactic {\it VI}$_c$ period-Wesenheit function (Fig.~\ref{fig-wf}), and to establish the Cepheid's position within the Centaurus arm (Fig.~\ref{fig-gs}).  TW Nor may be used as a high-weight calibrator for classical Cepheid relations.  Continued work on the Galactic calibration is needed.

\subsection*{{\rm \scriptsize ACKNOWLEDGEMENTS}}
\scriptsize{DJM is grateful to the following individuals and consortia whose efforts and surveys were the foundation of the study: 2MASS, P. Stetson (DAOPHOT), OGLE, the Araucaria project, V. Strai{\v z}ys, J-C. Mermilliod, F. van Leeuwen, F. Benedict, B. Skiff, L. Berdnikov, and staff at the CDS, arXiv, and NASA ADS.  We gratefully acknowledge use of data from the ESO Public Survey programme ID 179.B-2002, the Cambridge Astronomical Survey Unit, and funding from the FONDAP Center for Astrophysics 15010003, the BASAL CATA Center for Astrophysics and Associated Technologies PFB-06, the MILENIO Milky Way Millennium Nucleus from the Ministry of Economics ICM grant P07-021-F and Proyecto FONDECYT Regular 1090213. RS acknowledges financial support from CONICYT through GEMINI Project Nr. 32080016.}

\end{document}